\documentclass[11pt]{article}
\usepackage{times}

\newcommand{\arXiv}{{\sf arXiv}}
\newcommand{\mysection}[1]{\section{#1}}
\newcommand{\mysubsection}[1]{\subsection{#1}}
\newcommand{\myrules}{\rule{\textwidth}{0.3mm}\vspace*{0.5ex}\\}
\newcommand{\myrulee}{\vspace*{-2ex}\rule{\textwidth}{0.3mm}\vspace*{-3ex}\\}

\newcommand{\bul}{\item[$\bullet$]}

\newcommand{\lt}{$\rm <$}

\begin{document}

\pagestyle{empty}
\title{Open Archives Initiative protocol development\\ 
       and implementation at \arXiv}

\author{
Simeon Warner\\
(Los Alamos National Laboratory, USA)\\
({\tt simeon@lanl.gov})}

\date{23$^{\rm rd}$ January 2001}
\maketitle
\begin{abstract}
I outline the involvement of the Los Alamos e-print archive ({\arXiv})
within the Open Archives Initiative (OAI) and describe the implementation
of the {\em data provider} side of the OAI protocol v1.0. I highlight
the ways in which we map the existing structure of {\arXiv} onto 
elements of the protocol.
\footnote{Expanded version of talk presented at 
Open Archives Initiative Open Meeting in Washington, DC, USA
on 23$^{\rm rd}$ January 2001.}
\end{abstract}

\mysection{Introduction}
\mysubsection{Background}

The Los Alamos e-print archive, now called {\arXiv}~\cite{arXiv} and formerly
known as `xxx', was started by Paul Ginsparg in August 1991. 
It allows unrefereed author self-archiving of research papers and 
no-fee retrieval by users worldwide. 
Initially {\arXiv} was limited to high-energy theoretical physics, operated
over email only, and served $\approx$200 users. The number of users 
grew to over 1000 in a few months and has grown to over 70,000 now.
The scope of archive has expanded and new facilities have been 
added steadily since 1991~\cite{WhatsNew}. The following is a
list of some of the more significant developments:

\begin{list}{}{
\settowidth{\labelwidth}{from 1996}
\settowidth{\leftmargin}{from 1996xx}}
\item[Aug 1991] Physics e-print archive started:
  {\tt hep-th} archive with email interface.
\item[1992] ftp interface added.
  {\tt hep-ph} and {\tt hep-lat} added locally; 
  {\tt alg-geom}, {\tt astro-ph} and {\tt cond-mat} added remotely.
\item[Dec 1993] Web interface added.
\item[Nov 1994] Data at some remote archives (using the same software) 
  moved to main site, the remote sites become mirrors.
\item[Jun 1995] Automatic PostScript generation from {\TeX} source.
\item[Apr 1996] PDF generation added.
\item[Jun 1996] Web upload facility added.
\item[from 1996] Worldwide mirror network grows.
\item[from 1999] {\arXiv} involved in the OAI.
\end{list}

\mysubsection{The present}
%
% What is arXiv? Suggest significance
% What community does arXiv serve?
% Note speed of dissemination
% Note political mission
% Small database but quite heavily used, considered essential by
% some communities
% 

{\arXiv} is the dominant means of information dissemination within
certain areas of physics, notably high-energy physics, and has growing 
significance in many others~\cite{Blurb}. It has rendered the paper distribution of
pre-prints obsolete. The following is a list of some key statistics:

\begin{itemize}
\bul Covers physics, mathematics, computer science, and non-linear systems.
\bul Serves over 70,000 users in over 100 countries.
\bul Estimated 13 million downloads in 2000.
\bul Over 30,000 new submissions in 2000, over 150,000 e-prints total
  (approximately linear growth in submission rate, 
  $\approx$3500 extra each year).
% high-energy physics archives have 'saturated' to some extent and show
% very little change over the past three years. Other archives, such as
% astro-ph and cond-mat are growing rapidly
% 16,000 email listing subscribers (17Jan2001)
% allows revision but datestamps versions and keeps old versions
% use is free both for submission and reading
\bul $>$99\% of submissions entirely automated.
\bul Submission via web (68\%), email (27\%) and ftp (5\%).
% The e-print archive has certainly changed the means of information
% dissemination within some areas of physics. It has not yet led to
% changes in the peer-review system, though now this often occurs
% somewhat after a paper has been widely available and read. However,
% widespread archiving effectively separates the distribution and 
% peer-review functions and may lead to new models.
\bul Some journals now accept and {\arXiv} identifiers instead of requiring 
direct submission (e.g. APS: Phys. Rev. D, Elsevier: Phys. Lett. B).
\bul Los Alamos site funded by DOE\footnote{{\arXiv} has direct DOE grant 
funding and also support from the `Library Without Walls' project (LWW/STB-RL) 
within Los Alamos National Lab.} and NSF; mirror sites funded locally.
\end{itemize}

\mysubsection{{\arXiv} software}

The software running {\arXiv} comprises of the order of 30,000 lines of 
Perl running under Linux with numerous other programs 
({\TeX}, ghostscript, tar, gnuzip,...). 
The Perl code has evolved through 
many incremental changes on a system that has run continuously
for 9.5 years, punctuated only by moves to newer hardware and a few short
outages. We are currently putting significant effort into tidying and
rewriting the code in a modular fashion but, unfortunately, we do not
have the resources to engage in a complete off-line rewrite.
 
The growth of the {\arXiv} far beyond early expectations means that 
some aspects will need to be changed. 
For example, the metadata is currently stored in plain files and will 
probably be moved to an XML database.
This would allow more uniform parsing and character encoding, and
provide better extraction facilities for services such as the OAI
{\em data provider} interface. Also, the 3-digit serial number built into
our identifiers limits each archive to 999  submissions
per month (000 is not used). In November 2000, the {\tt astro-ph} 
archive had 583 submissions. 
This limitation makes it impossible to combine the current array of physics archives
into one archive with an extended subject-class structure as exists for other
subject areas on {\arXiv}.
All changes must be implemented without disrupting the
operation of the main site or the mirror sites.

\mysubsection{Involvement of {\arXiv} in the OAI}
The Open Archives Initiative (OAI) developed from a meeting held in Santa Fe 
in 1999 which was initiated by Paul Ginsparg ({\arXiv}, Los Alamos National Lab.), 
Rick Luce (Los Alamos National Lab.) and Herbert Van de Sompel
(University of Ghent, Los Alamos National Lab.). 
{\arXiv} has continued to be actively involved in both management 
of the initiative and technical development of the protocol. 

The protocol that resulted from the Santa Fe meeting~\cite{SantaFeMeeting}
was a subset of the Dienst protocol~\cite{Dienst} developed at
Cornell University. While the syntax has changed significantly, the
philosophy remains similar and our implementation has developed from
that written to comply with the Santa Fe Convention and announced
on 15$^{\rm th}$ February 2000. 

The initial focus of the OAI was author self-archived scholarly
literature --- e-prints, as they are often known. While the scope of the
OAI has expanded considerably, the e-print community has led the
protocol development. An example of this lead is the {\tt eprints.xsd} 
schema (appendix 2 in the protocol specification~\cite{OAIprotocol}) 
which defines an e-print community specific {\tt description} section 
for the response to the Identify verb.

\mysection{OAI protocol v1.0 implementation}
In the remainder of this paper I describe the {\arXiv} implementation of
the {\em data provider} side of the OAI protocol~v1.0~\cite{OAIprotocol}.
I discuss issues associated with some of the concepts used in
the protocol and then comment on each of the verbs.

\mysubsection{Identifiers}

Internally, the {\arXiv} software uses identifiers of the
form {\tt arch-ive/0101027} to refer to the latest version of an e-print,
where {\tt arch-ive} is the archive name (e.g. {\tt hep-th} or {\tt math},
and {\tt 0101027} is built up from last two digits of the year {\tt YY}, 
the month number {\tt MM}, and a serial number {\tt NNN}.
Some archives have compulsory subject-classes and 
the primary subject class is indicated by a two letter code
appended to the archive name giving a complete identifier of the
form {\tt arch-ive.SC/0101027}, where {\tt SC} is the subject-class.
The subject-class is not necessary in order to locate the e-print
but the form of the identifier including the subject-class is 
considered the canonical form.

There are multiple versions of some e-prints and the identifier just described is
considered to refer to the latest version.
A suffix {\tt vN} is used to denote a specific version, where N is 
the version number (starting at 1). The details of the internal identifiers are likely
to change to circumvent the limitations mentioned earlier.
The existing identifiers must, however, continue to work because
they have been widely used a references.

The OAI protocol specifies that identifiers must follow the URI~\cite{URI}
syntax. Our implementation exposes only the metadata of the latest version 
of each e-print and we follow the rules of the {\tt oai-identifier} type
shown in appendix 2 of the OAI protocol specification~\cite{OAIprotocol}.
We prepend to our internal identifier 
the scheme name ({\tt oai}), 
a colon separator ({\tt :}), 
our repository name ({\tt arXiv}), 
and another colon ({\tt :}). The resulting identifiers have the form
{\tt oai:arXiv:arch-ive[.SC]/0101027}. The following are example internal
identifiers and the corresponding OAI identifiers:

\begin{verbatim}
  hep-th/9901001        oai:arXiv:hep-th/9901001
  quant-ph/9912010      oai:arXiv:quant-ph/9912010
  math.SG/0001001       oai:arXiv:math.SG/0001001 
  cs.SE/0101002         oai:arXiv:cs.SE/0101002
\end{verbatim}

\mysubsection{Datestamps}

Until now, the significant dates in {\arXiv} records have been the dates
of submission of different revisions (as recorded in the {\tt Date} lines
of the metadata, and in the daily log file) and the dates of 
cross-listings (as recorded in the daily log file). We have had
no need to record by-hand modifications to the metadata or the
date of addition or modification of a journal-reference (which is the
only metadata element an author is allowed to change without generating
a new revision).

In the context of the OAI, we need a datestamp that reflects {\em any}
change in the metadata so that any change in the metadata will cause
the metadata to be re-harvested. 
We store the metadata for each e-print in a separate file, so the easiest 
way to provide the required datestamp is to use the file modification date
extracted from the file modification time.

It is straightforward to extract such a datestamp for a single record. 
However, the requirement for a search by datestamp
as used in the ListIdentifiers and ListRecords requests requires that
we compile indexes of datestamps. Further, the need to support incremental
harvesting without missing any changes means we need to rebuild (or at
least update) these indexes at whichever is the larger of the
update interval 
({\lt}1 day for by-hand modifications, 1 day for submissions, etc.)
or the datestamp resolution (1 day). If a longer interval were used then
it would be necessary to create datestamps tied to the rebuild times to
avoid a {\em service provider} that harvested more frequently (e.g. daily)
missing updates.

Within the existing protocol there is still the possibility for 
problems dependent upon the time of day the harvesting occurs in
relation to any updates. The simplest solution is for harvesters 
to overlap successive harvests by 1 day so that any updates occuring 
on the day of the previous harvest are sure to be included, even if they 
occured after the previous harvest. Double-harvesting a few records should
not cause any problems.

\mysubsection{Sets}

The OAI protocol characterizes sets as ``an optional construct for grouping 
items in a repository for the purpose of selective harvesting of records'',
and they may be arranged in {\em zero or more} hierarchies.
{\arXiv} has a two- or three-level (depending on subject area) grouping 
hierarchy based on the subject of the e-print. The three levels are:
\begin{description}
\item[group] There are four groups: physics, math, cs, nlin
\item[archive] The physics group has many archives (e.g. {\tt hep-th}, 
 {\tt astro-ph}, {\tt cond-mat} and even {\tt physics}). 
 Group and archive may be identified for the three other groups.
\item[subject-class] Some archives have subject-classes and individual e-prints
  may belong to one or more subject-class.
  However, we make a distinction
  between the {\em primary} subject-class and any others (which are treated
  in the same way as cross-lists to other archives).
\end{description}

It is possible to cross-list e-prints to archives other than the primary archive,
or other subject-classes within the same or other archives. 
The purpose of cross-lists is to avoid the need to submit a single e-print to
multiple archives. For example, if an e-print in the {\tt astro-ph} archive
is cross-listed to the {\tt hep-ph} archive, then `readers' of {\tt hep-ph} 
will see the {\tt astro-ph} e-print in the announcement email and the 
daily-listings on the web. The cross-listed {\tt astro-ph} e-print will also 
be included in searches of the {\tt hep-ph} archive.

We feel that it is desirable to reflect the existing {\arXiv} organization,
including the idea of cross-lists, in the OAI set structure. We currently
declare four sets which correspond to the four groups listed above. 
If an e-print in one group is cross-listed to another group, or an archive within
another group, then it will appear in the sets for both groups.
If there were a need for it then we could also expose the subject-class
layer in the set structure (skipping the archive layer which is an 
historical artifact). This was demonstrated in a prototype implementation.

Data from {\arXiv} has previously been harvested in a manner that would
now be achieved with sets. Data from the computer science archive
(now the {\tt cs} set) was harvested by the NCSTRL~\cite{NCSTRL} project
using a version of the Dienst protocol~\cite{Dienst}.

It is worth noting that {\em data providers} do not have to implement sets.
Also, even when harvesting from {\em data providers} that implement sets, 
{\em service providers} do not have to use the sets (i.e. they never specify
a {\tt set}).

\mysubsection{Deleted records}

{\arXiv} does not allow e-prints to be removed once submitted. We do
permit authors to submit a withdrawal notice as a new revision which
means that anyone looking for that e-print will see that the author
wishes for it to be considered withdrawn. However, earlier revisions
remain available. From the perspective of our OAI implementation,
we export metadata for the withdrawal notice in the same was as for 
any other e-print. 

There are a very small number (9 at present) of e-prints that have been 
removed because they were inappropriate or were duplicate copies. 
We have chosen to use the OAI {\tt status="deleted"} attribute for these 
papers. This is currently implemented using a lookup table.

\mysubsection{Metadata formats}

We disseminate metadata for e-prints in the following formats:
\begin{description}
\item[{\tt oai\_dc}] Dublin Core encoded in XML.
\item[{\tt oai\_rfc1807}] RFC1807 encoded in XML.
\item[{\tt arXiv}] Test-bed for new internal XML metadata format.
\item[{\tt arXivOld}] XML encoded version of current internal metadata format. 
\end{description}

Details of the {\tt oai\_dc} and {\tt oai\_rfc1807} formats are given by the
schemas in appendix 1 of the protocol specification~\cite{OAIprotocol}.

Many mappings are obvious, 
e.g. Title $\rightarrow$ Dublin Core `title', and 
     Abstract $\rightarrow$ Dublin Core `description'.
Other elements in Dublin Core and RFC1807 have to be extracted from
less specific fields in the current {\arXiv} metadata. One example
is the language of the e-print. Most e-prints on {\arXiv} are in 
English but some are in other languages (with an English abstract). The
language is usually reflected by a comment such as `in French' in the
our `Comments' field. We look for such declarations and export the
information in the `language' field of RFC1807 metadata.

We currently store author names and affiliations commingled in a rather 
free format which is quite difficult to untangle in general. We attempt
to extract the surnames, forenames or initials, surname prefixes 
(e.g. de, von), surname suffixes (e.g. III) and affiliations. 
For example, {\arXiv} metadata might say:
\begin{verbatim}
Authors: Fred A Bloggs, Mark Smith II (Univ A), T Sawyer (Univ B)
\end{verbatim}
where, following convention, we assume {\tt Fred A Bloggs}  has affiliation
{\tt Univ A\/} and we must recognize {\tt II} as a suffix rather than the
surname of {\tt Mark Smith}. 
Involvement in the OAI has highlighted the
need for {\arXiv} to collect better metadata.

\mysubsection{Document verb}

This verb is not part of the OAI protocol specification and the HTTP 
return is an error ({\tt 400 Malformed request}). 
However, an HTML page is also returned and this provides some notes
about the current status of our implementation and some links to
example requests for other verbs.
The request is:

\verb|http://arXiv.org/oai1?verb=Document| ,

where \verb|http://arXiv.org/oai1| is the BASE-URL of the {\arXiv}
OAI implementation, and \\
\verb|verb=Document| is the keyword argument specifying the verb.

\mysubsection{Identify verb}

Implementation of the Identify verb is simply a matter of generating an
appropriately formatted XML record containing information from
various configuration variables. The request and response are shown
in figure~\ref{Iverb}.
{\arXiv} returns two {\tt description} containers:

\begin{description}
\item[{\tt oai-identifier}]
Declares our use of the OAI identifier scheme defined by 
\verb|oai-identifier.xsd|, and gives a sample identifier 
\verb|oai:arXiv:quant-ph/9901001| .
\item[{\tt eprints}]
As we are part of the e-prints community, we include a description 
container that follows the {\tt eprints.xsd}
schema (appendix 2 of the protocol specification~\cite{OAIprotocol}).
\end{description}

\begin{figure}
\myrules
Request: \verb|http://arXiv.org/oai1?verb=Identify|

Response:
{\small \begin{verbatim}
<?xml version="1.0" encoding="UTF-8"?>
 <Identify xmlns="http://www.openarchives.org/OAI/1.0/OAI_Identify" 
   xmlns:xsi="http://www.w3.org/2000/10/XMLSchema-instance" 
   xsi:schemaLocation="http://www.openarchives.org/OAI/1.0/OAI_Identify 
                       http://www.openarchives.org/OAI/1.0/OAI_Identify.xsd">
  <responseDate>2001-01-22T10:01:27-07:00</responseDate>
  <requestURL>http://arXiv.org/oai1?verb=Identify</requestURL>
  <repositoryName>arXiv</repositoryName>
  <baseURL>http://arXiv.org/oai1</baseURL>
  <protocolVersion>1.0</protocolVersion>
  <adminEmail>local-admin@xxx.lanl.gov</adminEmail>
  <description>
   <oai-identifier xmlns="http://www.openarchives.org/OAI/oai-identifier" 
     xmlns:xsi="http://www.w3.org/2000/10/XMLSchema-instance" 
     xsi:schemaLocation="http://www.openarchives.org/OAI/oai-identifier 
                         http://www.openarchives.org/OAI/oai-identifier.xsd">
    <scheme>oai</scheme>
    <repositoryIdentifier>arXiv</repositoryIdentifier>
    <delimiter>:</delimiter>
    <sampleIdentifier>oai:arXiv:quant-ph%2F9901001</sampleIdentifier>
   </oai-identifier>
  </description>
  <description>
   <eprints xmlns="http://www.openarchives.org/OAI/eprints" 
     xmlns:xsi="http://www.w3.org/2000/10/XMLSchema-instance" 
     xsi:schemaLocation="http://www.openarchives.org/OAI/eprints 
                         http://www.openarchives.org/OAI/eprints.xsd">
    <content>
     <text>Author self-archived e-prints</text>
    </content>
    <metadataPolicy>
     <text>Metadata harvesting permitted through OAI interface</text>
     <URL>http://arXiv.org/help/oa/metadataPolicy</URL>
    </metadataPolicy>
    <dataPolicy>
     <text>Full-content harvesting not permitted 
           (except by special arrangement)</text>
     <URL>http://arXiv.org/help/oa/dataPolicy</URL>
    </dataPolicy>
    <submissionPolicy>
     <text>Author self-submission preferred, submissions screened 
           for appropriateness</text>
     <URL>http://arXiv.org/help/submit</URL>
    </submissionPolicy>
   </eprints>
  </description>
 </Identify>
\end{verbatim} }
\myrulee
\caption{\label{Iverb} Example Identify request and response}
\end{figure}

\mysubsection{ListSets verb}

This is very straightforward to implement.
The groups, the top level of our subject-classification structure, 
already have long and short names in the existing {\arXiv} code.
These names aretaken from configuration variables and written out as XML. 
The request and response are shown in figure~\ref{LSverb}.

\begin{figure}
\myrules
Request: \verb|http://arXiv.org/oai1?verb=ListSets|

Response:
{\small \begin{verbatim}
 <ListSets xmlns="http://www.openarchives.org/OAI/1.0/OAI_ListSets" 
   xmlns:xsi="http://www.w3.org/2000/10/XMLSchema-instance" 
   xsi:schemaLocation="http://www.openarchives.org/OAI/1.0/OAI_ListSets 
                       http://www.openarchives.org/OAI/1.0/OAI_ListSets.xsd">
  <responseDate>2001-01-22T10:02:48-07:00</responseDate>
  <requestURL>http://arXiv.org/oai1?verb=ListSets</requestURL>
  <set>
   <setSpec>nlin</setSpec>
   <setName>Nonlinear Sciences</setName>
  </set>
  <set>
   <setSpec>math</setSpec>
   <setName>Mathematics</setName>
  </set>
  <set>
   <setSpec>physics</setSpec>
   <setName>Physics</setName>
  </set>
  <set>
   <setSpec>cs</setSpec>
   <setName>Computer Science</setName>
  </set>
 </ListSets>
\end{verbatim} }
\myrulee
\caption{\label{LSverb} Example ListSets request and response.}
\end{figure}

As we have a small number of sets there has been no need to implement
the partial response and acceptance of a {\tt resumptionToken}. Any
request which supplies a {\tt resumptionToken} will result in
a 400 error.

\mysubsection{ListMetadataFormats verb}

The metadata formats that can be disseminated by the {\arXiv} OAI interface
are stored in configuration variables (along with pointers to the routines
that convert from our native format). Implementation of this verb is simply
a matter of writing out this information in the correct format.
The request and response are shown in figure~\ref{LMFverb}.

\begin{figure}
\myrules
Request: \verb|http://arXiv.org/oai1?verb=ListMetadataFormats|

Response:
{\small \begin{verbatim}
<?xml version="1.0" encoding="UTF-8"?>
 <ListMetadataFormats 
   xmlns="http://www.openarchives.org/OAI/1.0/OAI_ListMetadataFormats" 
   xmlns:xsi="http://www.w3.org/2000/10/XMLSchema-instance" 
   xsi:schemaLocation=
"http://www.openarchives.org/OAI/1.0/OAI_ListMetadataFormats 
 http://www.openarchives.org/OAI/1.0/OAI_ListMetadataFormats.xsd">
  <responseDate>2001-01-25T16:30:23-07:00</responseDate>
  <requestURL>http://arXiv.org/oai1?verb=ListMetadataFormats</requestURL>
  <metadataFormat>
   <metadataPrefix>arXivOld</metadataPrefix>
   <schema>http://arXiv.org/OAI/arXivOld.xsd</schema>
   <metadataNamespace>http://arXiv.org/OAI/</metadataNamespace>
  </metadataFormat>
  <metadataFormat>
   <metadataPrefix>arXiv</metadataPrefix>
   <schema>http://arXiv.org/OAI/arXiv.xsd</schema>
   <metadataNamespace>http://arXiv.org/OAI/</metadataNamespace>
  </metadataFormat>
  <metadataFormat>
   <metadataPrefix>oai_rfc1807</metadataPrefix>
   <schema>http://www.openarchives.org/OAI/rfc1807.xsd</schema>
   <metadataNamespace>
    http://info.internet.isi.edu:80/in-notes/rfc/files/rfc1807.txt
   </metadataNamespace>
  </metadataFormat>
  <metadataFormat>
   <metadataPrefix>oai_dc</metadataPrefix>
   <schema>http://www.openarchives.org/OAI/dc.xsd</schema>
   <metadataNamespace>http://purl.org/dc/elements/1.1/</metadataNamespace>
  </metadataFormat>
 </ListMetadataFormats>
\end{verbatim} }
\myrulee
\caption{\label{LMFverb} Example ListMetadataFormats request and response.}
\end{figure}

As we have a small number of metadata formats there has been no need 
to implement the partial response and acceptance of a {\tt resumptionToken}.
Any request which supplies a {\tt resumptionToken} will result in
a 400 error.

\clearpage
\mysubsection{GetRecord verb}

The majority of the effort involved in implementing the GetRecord
verb is in performing the metadata format conversion and mapping
from our internal format to the format requested. 
This has been discussed above. Once the request has been parsed, there are
four possible outcomes:
\begin{enumerate}
\item Item does not exist $\rightarrow$ no {\verb|<record>|} container returned.
\item Item is `deleted' $\rightarrow$ {\verb|<record status="deleted">|} container 
  with {\verb|<header>|} block returned.
\item Item exists but can not be disseminated in the requested metadata format 
  $\rightarrow$ {\verb|<record>|} container with {\verb|<header>|} but 
  no {\verb|<metadata>|} block returned.
\item Item exists and can be disseminated in the requested metadata format 
  $\rightarrow$ {\verb|<record>|} container with {\verb|<header>|} and 
  {\verb|<metadata>|} blocks returned.
\end{enumerate}

As metadata for all items in {\arXiv} can be disseminated in all four
metadata formats supported, case 3 applies only if an unsupported
metadata format is requested. Figure~\ref{GRverb} shows an example
GetRecord request and response.

We have chosen not to implement the {\verb|<about>|} container
at this time because we do not have rights information for individual
metadata records and consider that the overall statement supplied in
the Identify response is adequate.

\begin{figure}
\myrules
Request: \verb|http://arXiv.org/oai1?verb=GetRecord&|\\
\verb|  identifier=oai:arXiv:cs.DL/0101027&metadataPrefix=oai_dc|

Response:
{\small \begin{verbatim}
<?xml version="1.0" encoding="UTF-8"?>
 <GetRecord xmlns="http://www.openarchives.org/OAI/1.0/OAI_GetRecord"
   xmlns:xsi="http://www.w3.org/2000/10/XMLSchema-instance" 
   xsi:schemaLocation="http://www.openarchives.org/OAI/1.0/OAI_GetRecord 
                        http://www.openarchives.org/OAI/1.0/OAI_GetRecord.xsd">
  <responseDate>2001-01-26T09:41:13-07:00</responseDate>
  <requestURL>http://arXiv.org/oai1?verb=GetRecord&amp;
identifier=oai%3AarXiv%3Acs.DL%2F0101027&amp;metadataPrefix=oai_dc</requestURL>
  <record>
   <header>
    <identifier>oai:arXiv:cs.DL/0101027</identifier>
    <datestamp>2001-01-25</datestamp>
   </header>
   <metadata>
    <oai_dc xmlns="http://purl.org/dc/elements/1.1/" 
      xmlns:xsi="http://www.w3.org/2000/10/XMLSchema-instance" 
      xsi:schemaLocation="http://purl.org/dc/elements/1.1/ 
                           http://www.openarchives.org/OAI/dc.xsd">
     <title>Open Archives Initiative protocol development and implementation 
at arXiv</title>
     <creator>Warner, Simeon</creator>
     <subject>Digital Libraries</subject>
     <description>  I outline the involvement of the Los Alamos e-print 
archive (arXiv) within the Open Archives Initiative (OAI) and describe the 
implementation of the data provider side of the OAI protocol v1.0. I 
highlight the ways in which we map the existing structure of arXiv onto 
elements of the protocol.</description>
     <description>Comment: 15 pages. Expanded version of talk presented at 
Open Archives Initiative Open Meeting in Washington, DC, USA on 
23 January 2001</description>
     <date>2001-01-25</date>
     <type>e-print</type>
     <identifier>http://arXiv.org/abs/cs.DL/0101027</identifier>
    </oai_dc>
   </metadata>
  </record>
 </GetRecord>
\end{verbatim} }
\myrulee
\caption{\label{GRverb} Example GetRecord request and response.}
\end{figure}

\mysubsection{ListIdentifiers verb}

This verb is essentially a search by datestamp, the optional {\tt from}
and {\tt until} parameters specifying the datestamp range, and the optional 
{\tt set} parameter limiting the archives searched. 
We do not
want to return all $>$150,000 identifiers in one response and so implement
partial responses and supply a {\tt resumptionToken} as necessary. Our index
of updates is time ordered so we {\em choose} to return identifiers in
datestamp order and build the {\tt resumptionToken} from a concatenation
of the new {\tt from} parameter, the {\tt until} parameter and the
{\tt set}. This means that the {\tt resumptionToken} does not
expire at any particular time although the response will change if
the index of updates changes. 
An example ListIdentifiers request and response is shown in 
figure~\ref{LIverb}.

\begin{figure}
\myrules
Request: \verb|http://arXiv.org/oai1?verb=ListIdentifiers|

Response:
{\small \begin{verbatim}
<?xml version="1.0" encoding="UTF-8"?>
 <ListIdentifiers xmlns="http://www.openarchives.org/OAI/1.0/OAI_ListIdentifiers" 
   xmlns:xsi="http://www.w3.org/2000/10/XMLSchema-instance" 
   xsi:schemaLocation="http://www.openarchives.org/OAI/1.0/OAI_ListIdentifiers 
   http://www.openarchives.org/OAI/1.0/OAI_ListIdentifiers.xsd">
  <responseDate>2001-01-22T10:06:19-07:00</responseDate>
  <requestURL>http://arXiv.org/oai1?verb=ListIdentifiers</requestURL>
  <identifier>oai:arXiv:math.DS/9204240</identifier>
  <identifier>oai:arXiv:math.DS/9204241</identifier>
  <identifier>oai:arXiv:math.LO/9201250</identifier>
  ...
  <identifier>oai:arXiv:alg-geom/9202008</identifier>
  <identifier>oai:arXiv:alg-geom/9202009</identifier>
  <resumptionToken>1992-05-01___</resumptionToken>
 </ListIdentifiers>

\end{verbatim} }
\myrulee
\caption{\label{LIverb} Example ListIdentifiers request and response.}
\end{figure}

\mysubsection{ListRecords verb}

This verb is essentially a combination of the ListIdentifiers and 
GetRecord verbs. We implement the search by datestamp and set using the 
same code as for ListIdentifiers. Then, for each identifier found, we use
the same code the implements the GetRecord verb to write XML
metadata records in the format requested.
An example request and response is shown in figure~\ref{LRverb}.

\begin{figure}
\myrules
Request: \verb|http://arXiv.org/oai1?verb=ListRecords&metadataPrefix=oai_dc|

Response:
{\small \begin{verbatim}
<?xml version="1.0" encoding="UTF-8"?>
 <ListRecords xmlns="http://www.openarchives.org/OAI/1.0/OAI_ListRecords" 
   xmlns:xsi="http://www.w3.org/2000/10/XMLSchema-instance" 
   xsi:schemaLocation="http://www.openarchives.org/OAI/1.0/OAI_ListRecords 
                       http://www.openarchives.org/OAI/1.0/OAI_ListRecords.xsd">
  <responseDate>2001-01-22T10:08:02-07:00</responseDate>
  <requestURL>http://arXiv.org/oai1?verb=ListRecords&amp;
metadataPrefix=oai_dc</requestURL>
  <record>
   <header>
    <identifier>oai:arXiv:math.DS/9204240</identifier>
    <datestamp>1992-04-01</datestamp>
   </header>
   <metadata>
    <oai_dc xmlns="http://purl.org/dc/elements/1.1/" 
      xmlns:xsi="http://www.w3.org/2000/10/XMLSchema-instance" 
      xsi:schemaLocation="http://purl.org/dc/elements/1.1/ 
                          http://www.openarchives.org/OAI/dc.xsd">
     <title>Dynamics of certain non-conformal semigroups</title>
     <creator>Jiang, Yunping</creator>
     <subject>Dynamical Systems</subject>
     <description>  A semigroup generated by two dimensional $C^{1+\alpha}$ 
contracting maps is considered. 
     ...
     </description>
     <date>1992-04-01</date>
     <type>e-print</type>
     <identifier>http://arXiv.org/abs/math.DS/9204240</identifier>
    </oai_dc>
   </metadata>
  </record>
  <record>
   <header>
    <identifier>oai:arXiv:math.DS/9204241</identifier>
    <datestamp>1992-04-20</datestamp>
   </header>
   <metadata>
   ...
   </metadata>
  </record>
  ...
  <resumptionToken>1992-05-01___dc</resumptionToken>
 </ListRecords>
\end{verbatim} }
\myrulee
\caption{\label{LRverb} Example ListRecords request and response.}
\end{figure}

\mysubsection{Flow Control}

The main {\arXiv} site, {\tt http://arXiv.org/}, is heavily used
and has various automated scripts to prevent badly written and
non-conforming (i.e. not obeying {\tt /robots.txt}) robots from
loading the server to the point where there is denial-of-service
to other users. {\arXiv} is particularly vulnerable because of
the fact that most papers are stored as {\TeX} source and 
processed to produce PostScript or PDF on demand (with a large
cache).
Flow control is thus essential to avoid legitimate OAI 
{\em service providers} getting blocked.

Flow control is implemented with a combination of HTTP 503 Retry-After
replies and, for the list requests, the use of partial responses
and the {\tt resumptionToken}. A harvester that abides by the
delay specified in the Retry-After reply will not get blocked.
The strategy is simple: the OAI code keeps track of the time of the
last request from each site and enforces a minimum before answering
further requests. If a request 
comes sooner than is permitted, a Retry-After reply is issued 
where the delay returned is the time that the harvester must wait
before a request will be answered. To avoid loading the server
with searches by date, there is a longer delay for ListIdentifiers
and ListRecords requests than for other requests. A harvester that
knew the delay could pre-emptively wait between requests. With the current
protocol there is no way to communicate this delay so the usual situation
will be that every fulfilled request is followed by a request answered
by a Retry-After reply indicating the time the harvester must wait. 

Before the addition of flow control we had repeated problems with 
OAI developers getting blocked by {\arXiv}'s robot detection scripts.
This is no longer a problem and the {\sf arc}~\cite{ARC} 
harvester has managed to harvest metadata for all {\arXiv} records
several times without running foul of the robot detection scripts.

One could also implement load-balancing over a number of servers
using the HTTP 302 response permitted by the OAI protocol. 
This has not been implemented for {\arXiv) as we currently run only
one server at the main site.

\mysubsection{Implementation cost}
The current implementation has evolved from the Open Archives 
Dienst Subset~\cite{Dienst} implementation so it is hard to estimate 
the time it would take to write the interface from scratch using the OAI
v1.0 protocol specification. The {\arXiv} implementation is complicated
significantly by internal inconsistencies in the archive structure.
The list below indicates some of the major efforts:
\begin{description}
\item[`Santa Fe convention', OAI Dienst Subset, December 1999]{
  $\approx$2-3 weeks effort which included time to understand Dienst, writing 
  utility routines such as an XML writing library (now available in
  standard libraries), coding a search by date, and coding conversion of
  our metadata to RFC1807.}
\item[OAI v0.2, October 2000]{
  $\approx$2 days which included writing \TeX$\rightarrow$UTF-8 
  conversion code for the special characters which are currently mostly 
  {\TeX} encoded in our metadata. Also included rewriting parsing code 
  for simplified syntax.}
\item[OAI v1.0, January 2001]{
  Many small changes during protocol development, most of which involved
  simplifications in the code! 
}
\end{description}
The majority of the effort necessary to create a new implementation is
likely to be in routines to implement
metadata format conversion; 
a search to find records by datestamp; and
perhaps flow control through partial responses and Retry-After returns.

\mysection{The future}
%
% What are we going to do?
% - work on e-print specific metadata format
% - need to establish consensus within the e-print community
% - see how use of <about> and rights part of protocol develops
%
% What do we hope others will do?
% - resource discovery tools, increase visibility of arXiv
%
I hope that the OAI protocol will be widely adopted as a means of
exposing metadata by the scholarly publishing community and by others.
The utility of this is contingent upon the development services using 
this metadata.
One can imagine resource discovery tools that index all OAI compliant 
repositories and using just the required Dublin Core~\cite{DC} metadata 
still provide much more structured search facilities than generic
Web search engines such as Google~\cite{Google}.

I am confident that the scholarly publishing community, especially the
e-print community, will be quick to use the OAI protocol.
There already exists one OAI-based cross-archive search engine 
({\sf arc}~\cite{ARC}). 
The e-prints community is also developing a metadata set
able to encode a richer set of information than Dublin Core
and to include relations between records. I hope that this
will lead to community specific tools with additional
functionality based on this extra information.

\mysection{Acknowledgements}

This work has been carried out as part of the development of {\arXiv}
in collaboration with Paul Ginsparg and Thorsten Schwander. I
am also thankful for many helpful discussions with other members of
the OAI Technical Working Group and the OAI v1.0 Alpha Test Group.

\end{document}